# Universal Non-Polar Switching in Carbon-Doped Transition Metal Oxides (TMOs) and Post TMOs

C.A. Paz de Araujo,[1, 2] Jolanta Celinska,[2] Chris R. McWilliams,[2] Lucian Shifren,[3] Greg Yeric,[3] X. M. Henry Huang,[3] Saurabh Vinayak Suryavanshi,[3] Glen Rosendale,[3] Valeri Afanas'ev,[4] Eduardo C. Marino,[5] Dushyant Madhav Narayan,[6] Daniel S Dessau[6]

1) *University of Colorado, Colorado Springs, Colorado, 80918*
2) *Symetrix Corporation, Colorado Springs, Colorado, 80919*
3) *Cerfe Labs, Austin, Texas, 78737*
4) *Katholieke Universiteit Leuven, 3000 Leuven, Belgium*
5) *Federal University of Rio de Janeiro, Rio de Janeiro, 21941-901, Brazil*
6) *Center for Experiments on Quantum Materials & Department of Physics, University of Colorado, Boulder, Colorado, 80309*

**ABSTRACT**

Transition metal oxides (TMOs) and post-TMOs (PTMOs), when doped with Carbon, show non-volatile current-voltage (I-V) characteristics, which are both universal and repeatable. We have shown spectroscopic evidence of the introduction of carbon-based impurity states inside the existing larger bandgap effectively creating a smaller bandgap which we suggest could enable Mott-like correlation effect. Our findings indicate new insights for yet to be understood unipolar and nonpolar resistive switching in the TMOs and PTMOs. We have shown that device switching is not thermal-energy dependent and have developed an electronic-dominated switching model that allows for the extreme temperature operation (from 1.5 K to 423 K) and state retention up to 673 K for a 1-hour bake. Importantly, we have optimized the technology in an industrial process and demonstrated integrated 1-transistor/1-resistor (1T1R) arrays up to 1 kbit with 47 nm devices on 300 mm wafers for advanced node CMOS-compatible correlated electron RAM (CeRAM). These devices are shown to operate with 2 ns write pulses and retain the memory states up to 200 $^0$C for 24 hours. The collection of attributes shown, including scalability to state-of-the-art dimensions, non-volatile operation to extreme low and high temperatures, fast write, and reduced stochasticity as compared to filamentary memories such as ReRAMs show the potential for a highly capable two-terminal back-end-of-line non-volatile memory.

1. Introduction

For well over a decade, metal-insulator transition (MIT) research has sought to implement resistive switching in a variety of ways [1]. Such efforts have led to ionic and vacancy reactions, which are generally accompanied by the electrochemical formation of filaments that act as the main switching mechanisms in TMO-based resistive random-access memories (ReRAMs) and conductive bridge random-access memories (CBRAMs) [2].

In this work, we report results in which carbon doping of TMOs and post-TMOs causes a creation of new impurity states which we suggest are responsible for the non-polar switching seen in our devices. We propose that the resulting universality of the current-voltage (I-V) characteristics is akin to the classic current electron density-driven MIT [3] which is a dramatic departure from ReRAMs and CBRAMs sometimes made with similar oxides. We identify these devices as CeRAM, reflecting the correlated electrons (Ce) that are suggested to participate in the electronic switching [4] [5]. We will show experimental data that indicate a distinct difference between MIT switching in TMOs and PTMOs with and without carbon and argue we have identified the key difference in transport that occurs in the high resistance state (HRS). Building on our previous work [4] [5] and new experimental data, we show a new understanding of the band structure in our devices.

We have shown a much wider temperature range of operation from 1.5 K to 423 K, which is very unique for non-volatile memory devices. We also discuss replication of similar behavior in multiple other materials including $HfO_2$, $Bi_2O_3$, $YrTiO_3$, and $PbNiO_3$, therefore, indicating the universality of the underlying phenomenon. Our control of the device characteristics by controlling carbon allows us to further engineer the device for the ultimate application. Towards that end, we integrate our process in a standard CMOS process on 300 mm wafers and make sub-50 nm CeRAM devices, which can operate with 2 ns wide pulses for both SET and RESET.

## 2. CeRAM Characteristics

Correlated electron random access memory (CeRAM) is a device realized by a metal-insulator-metal capacitor (MIM) where carbon-doped metal oxide is sandwiched between two metal electrodes as shown in Figure 1a. Our fabrication is accomplished on 4-inch wafers with spin-on deposition of carbon-doped NiO (unless stated otherwise) and a picture of a final wafer is shown in Figure 1b. The metal electrodes used in the study are platinum and we have also seen similar behavior with iridium electrodes. We later use physical vapor deposition (PVD) for metal oxide deposition on 300 mm wafers for the industrial fabrication process discussed in Section 5. Unless otherwise stated, each measurement is performed on a new virgin device, to make sure the device under test is not damaged from the previous test.

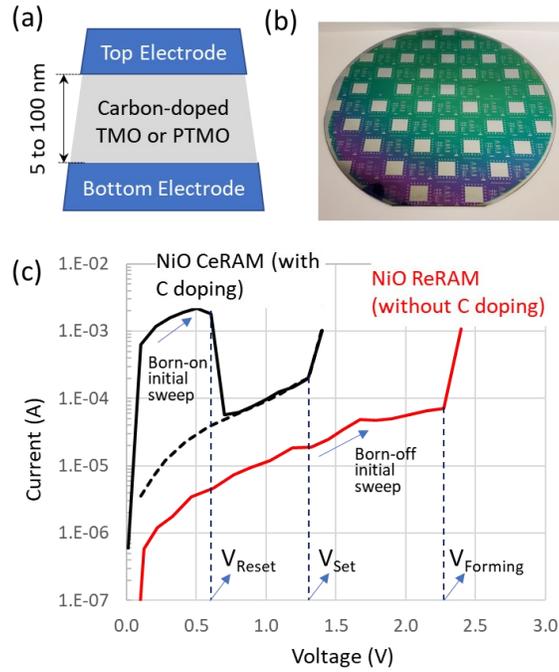

Figure 1: (a) Schematic of a CeRAM capacitor with metal-insulator-metal structure (b) 4-inch carbon doped NiO wafer with micron-size CeRAM structures (5 × 5 μm$^2$ devices). (c) The I-V characteristics of two different samples is shown: CeRAM when NiO is doped with carbon (black) is compared with ReRAM when NiO does not have carbon doping (Red).

The typical I-V characteristics of CeRAMs are shown in Figure 1c as the black curves. Each CeRAM curve is obtained by performing an increasing DC voltage sweep starting from 0 V with a dwell time of 4.5 ms per data point. Note that, unlike other resistive memory, we use voltage as input and read current as the output. All operations are performed in the direction of increasing voltage magnitude. The initial conduction (virgin state) is indicative of a metal-like phase (low resistance) that is "born-on" and there is no forming required. The low resistance state (LRS) of the virgin device switches to a semi-insulating high resistance state (HRS) around $V_{Reset}$ (RESET operation, defined in Figure 1c). If $V_{Reset}$ is not applied the non-volatile metal-like state does not change. Once transitioned to high resistance state, the CeRAM stays in a non-volatile state and only changes to metal-like state when $V_{Set}$ (SET operation, defined in Figure 1c) is applied. The change from semi-insulating to metal-like state is drastic and we apply an external current compliance to avoid device burn-out. The relatively constant current plateau from $V_{Reset}$ to $V_{Set}$ represents the reference semi-insulating phase of CeRAM. The SET and RESET sweeps are done alternatingly as shown in Figure 1c and represent one complete write cycle.

The programming window ( $|Vreset - Vset|$ ) is wide enough to allow each write operation securely. It is important to note that without carbon, we do not see CeRAM behavior and as shown with red solid line in Figure 1c, the device without any carbon doping is insulating until a breakdown or forming event occurs at higher voltage. Such characteristics are commonly observed in filamentary memory such as ReRAMs, where the initial operation is at low current and requires high switching voltage ($V_{Forming}$ as defined in Figure 1c). We note that the SET operation of CeRAM is different from ReRAM forming in terms of the magnitude of switching voltage ($V_{Set}$ vs. $V_{Forming}$) and leakage current. We further expand on the differences at the end of Sections 2 and 3.

**Non-polar switching:** When the voltage is reversed in polarity, a similar I-V signature is repeated in a perfectly symmetric manner. Both CeRAM SET and RESET are non-polar, as switching can be induced in either the positive or negative I-V quadrants, independent of the previous SET or RESET operation. In Figure 2, we show RESET and SETs with both positive and negative polarities. This is significantly different than the unipolar switching that is commonly observed and discussed in ReRAM literature [6], where both SET and RESET happen in single (either Q1 or Q3) quadrant but not in different quadrants as shown in the Figure 2.

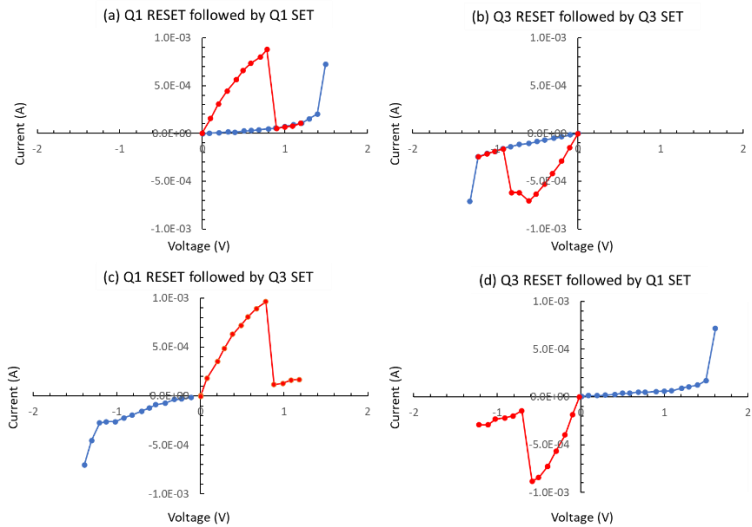

Figure 2: A non-polar switching in CeRAM. (a) Q1 RESET is followed by Q1 SET. (b) Q3 RESET is followed by Q3 SET. (c) Q1 RESET is followed by Q3 SET. (d) Q3 RESET is followed by Q1 SET.

**Critical voltage and current density**: In Figure 3a, we show a one-transistor one-resistor (1T1R) bitcell. The gate voltage (Vg) of the transistor controls the maximum current flowing through the transistor and, therefore, acts as a current compliance for the CeRAM device.

The transistor is used to artificially limit the current in RESET and SET operation. In Figure 3b, we first test the transistor without any CeRAM in series to test the maximum current that the transistor can carry at Vg = 0.9 V and Vg =0.95 V. The transistor only current-voltage curves are shown as dashed lines for reference. During the RESET for the solid blue curve, we artificially limit the current by applying Vg=0.9 V to the gate of the transistor in series with CeRAM (Figure 3b) and as a result, the CeRAM device does not RESET. In the case of the solid orange curve, we increase the transistor gate voltage to 0.95 V and can switch the CeRAM device. This indicates that if sufficient current is not supplied the CeRAM device will not RESET.

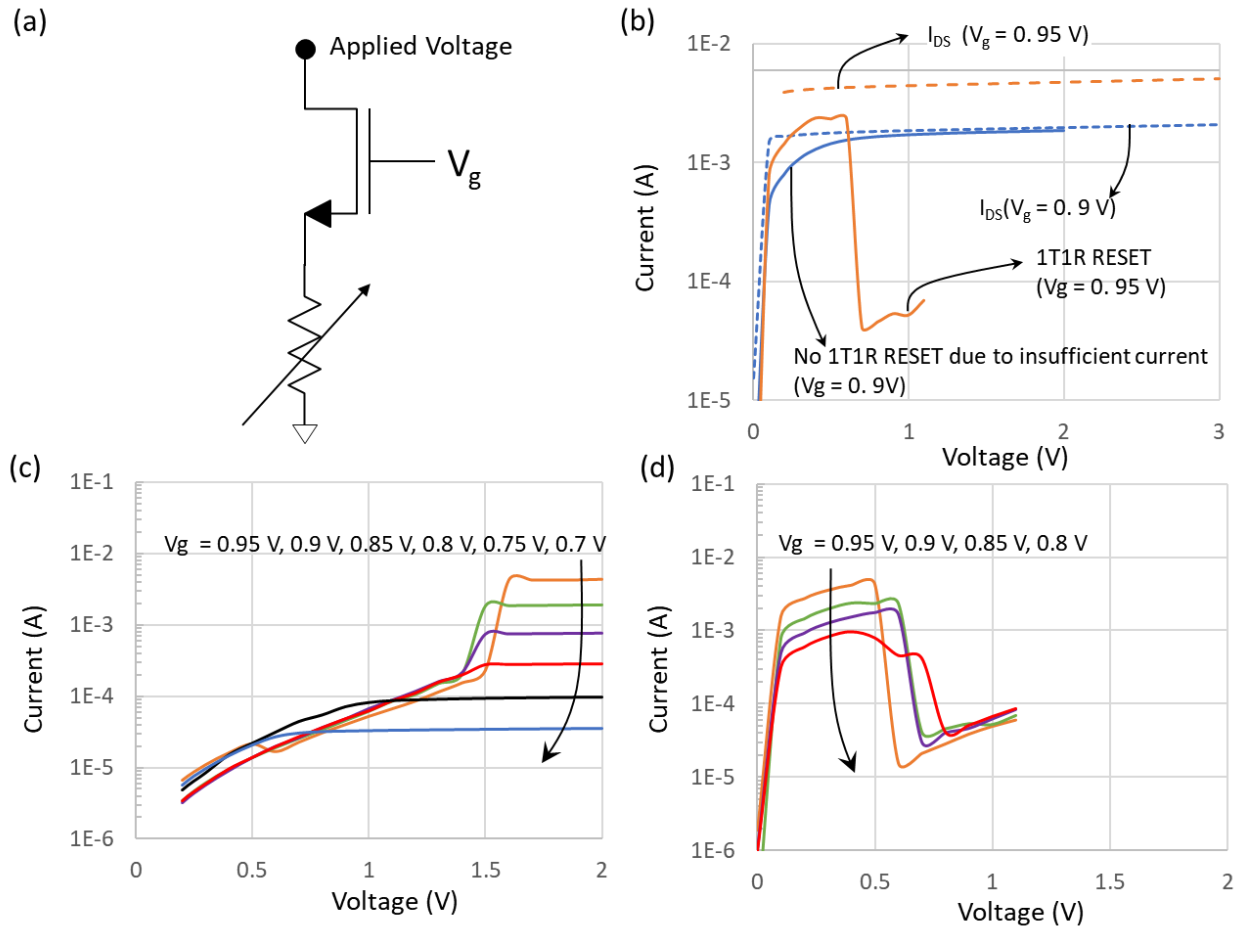

Figure 3: (a) Schematic diagram of a 1T1R circuit used to operate CeRAM devices. (b) Current characteristics of the selector transistor alone (dashed lines) and the select transistor in series with the CeRAM cell (solid lines). With $V_G$ = 0.9 V, the compliance current is set at 2 mA which increases to 5 mA on increasing $V_G$ to 0.95 V. Since the CeRAM RESET current for these devices is higher than 2 mA it only resets when the transistor gate voltage is 0.95 V and the compliance current is 5 mA. (c) SET I-V characteristics vs. applied voltage for various gate voltage of the access transistor. (d) Resulting RESET curves (post set) are shown by matching color.

In Figure 3c, we see a similar need for a sufficient current during the SET operation. If we reduce the transistor gate voltage below 0.75V, the CeRAM device does not undergo a SET operation. It is worth noting that depending on the compliance current during successful SET operations, the CeRAM device gets programmed in a new resistance state. For each successful SET, we show the resulting RESET operation in Figure 3d. Corresponding operations are color-matched: for example, the orange curve (with Vg = 0.95 V) in Figure 3c is immediately followed by the orange curve (with Vg = 0.95 V) in Figure 3d. As the compliance (SET) current in Figure 3c is reduced, we notice that the RESET current in Figure 3d is also reduced proportionally. This capability can be exploited to increase logical bit density and can also be used in analog neuromorphic computing [7].

For Vg of 0.75 V and 0.7 V, the transistor does not supply enough SET current and as a result the device remains in high resistance state. Note that in Figure 3c and Figure 3d., the SET and RESET voltages remain constant and do not change with the switching current. As long as both critical voltage and critical current are met, the CeRAM device will undergo switching. We expect this property to manifest as strong immunity to radiation, which is a critical problem for applications in space, military, and nuclear application.

**Cryogenic operation and metallic behavior**: Figure 4a shows non-polar (Q1 and Q3) CeRAM operation down to 1.5 K. We extract the RESET power during the operation and plot it as a function of temperature in Figure 4b. The RESET power of CeRAM reduces with ambient temperature and is lowest for 1.5 K. If the switching were thermally driven, the lower temperature would need higher input power to switch. In contrast, we see that the switching in CeRAM requires lower power as we reduce the ambient temperature and show that thermal energy plays an insignificant role in CeRAM operation.

Figure 4c shows the resistance state of the virgin devices as a function of the temperature, measured at a very low voltage to avoid self-heating or other effects. The increase in resistance as a function of temperature indicates metallic behavior which is not expected in insulating oxides. The weak metallic behavior of the virgin ON state helps us to define band structure of the carbon doped TMOs discussed later. Note that each plot has been taken on a virgin device to avoid additional noise from device-to-device variation.

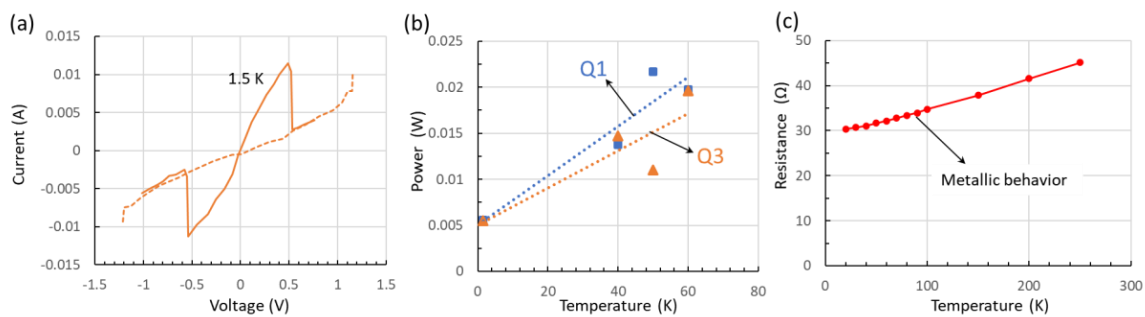

Figure 4: Cryogenic operation of CeRAM: (a) We show non-polar switching in Q1 and Q3 at 1.5 K. The lowest temperature is the limit of our test set up and is not a fundamental device limit. (b) The extracted reset power as a function of temperature for Q1 and Q3. (c) The resistance of the virgin device (before switching) as a function of temperature indicating weak metallic conduction.

**High temperature operation and retention**: We further test the CeRAM at high temperature extremes and Figure 5a shows non-polar operation at 423 K. This temperature is limited by our test set up and is not a fundamental limit of the devices. In addition to the operation, we also bake CeRAM devices after programming for 1-hour at various temperatures up to 673 K. As seen from the Figure 5b, the resistance of the devices does not change indicating strong non-volatile nature of CeRAM switching. The state retention at 673 K coupled with a wide operation range from 1.5 K to 423 K shows unique nature of the CeRAM switching mechanism and hints at the possibility of a non-thermal switching mechanism unlike the filamentary ReRAM and CBRAM devices, which show strong temperature dependence and are dependent on physical movements of atoms or vacancies.

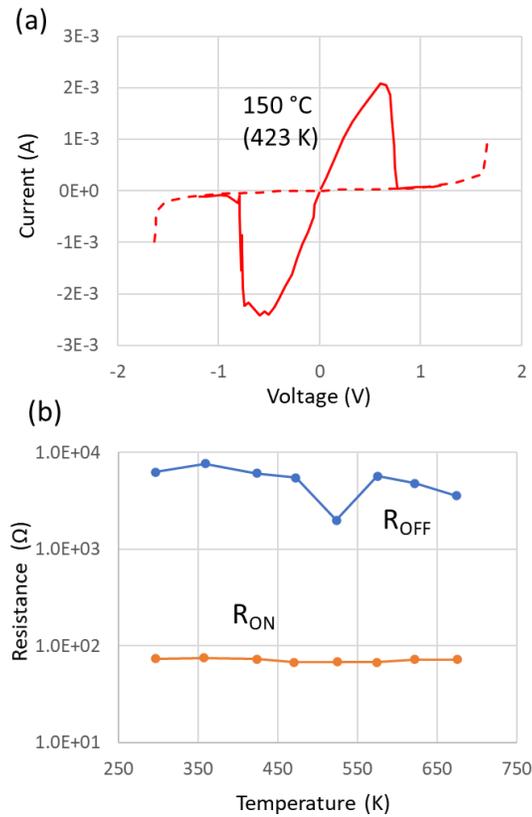

Figure 5: High temperature operations of CeRAM: (a) Device operation at 150 °C (423 K). The highest operation temperature is a limit of our test setup and not a fundamental device limit. (b) Devices set in ON and OFF state and baked at increasing temperature for one hour. The devices are cooled and measured at room temperature. Both states show no significant change in the resistance indicating strong non-volatile nature of CeRAM.

3. **Carbon doping**

Figure 6 shows the I-V curves for different carbon-doped oxides represented by unique colors for NiO, $PbNiO_3$, $HfO_2$, $YTiO_3$, and $Bi_2O_3$. This set of materials exhibiting CeRAM behavior includes TMOs and post-TMOs. The targeted carbon doping in these materials is 1%. However, measuring carbon concentration in the deposited films at such low carbon percentage is very difficult due to low atomic mass. We are actively investigating to accurately quantify the carbon doping in these films.

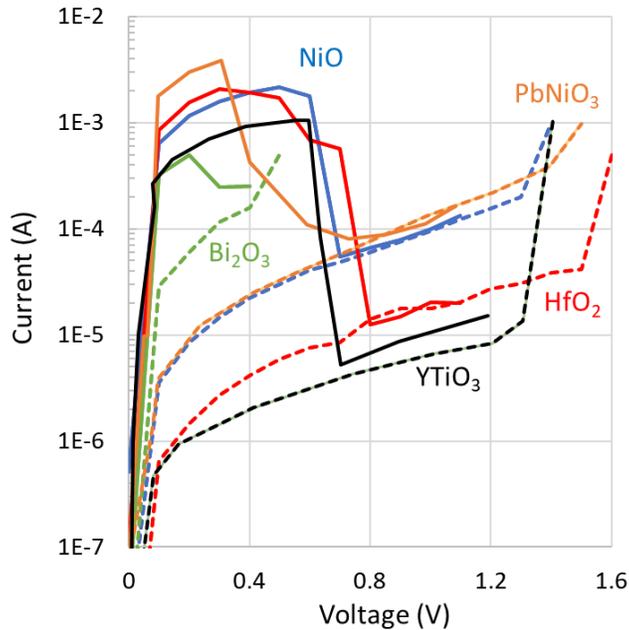

Figure 6: Examples of universal switching I-V (reset and set) in TMOs and post-TMO. The switching curves shown here are for NiO, PbNiO$_3$, HfO$_2$, Bi$_2$O$_3$, and YTiO$_3$. The bold lines represent the RESET curves, while the dashed lines represent the SET curves.

he solid lines are the initial sweep which shows the CeRAM devices starting in a low resistance metal-like state followed by RESET to a high resistance state. A subsequent sweep in the high resistance state shown by dashed lines, which terminate with a compliance current-limited SET operation back into a low resistance state. The similarities of these curves across this set of materials display the universality of the effect of carbon doping. As discussed in Figure 1, without the carbon doping none of the oxides will switch without first electro-forming. The different levels of current and voltage in different materials is beyond the scope of this paper but suggest the switching current and voltages across the material systems depends on the metal ion and carbon interaction.

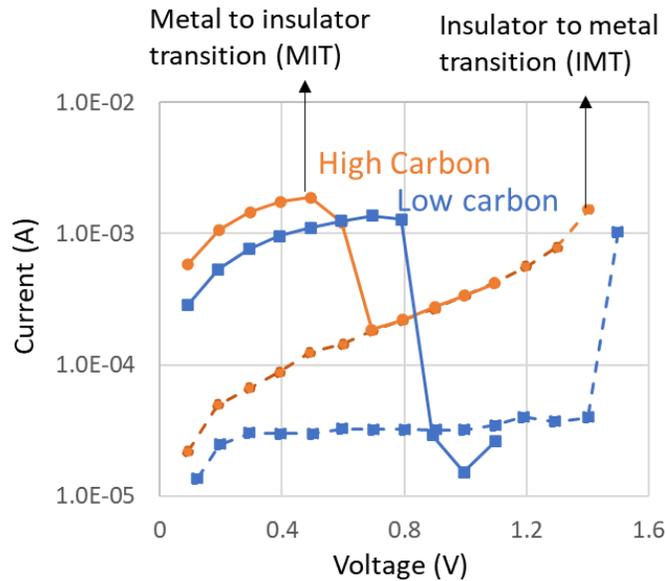

Figure 7: The I-V plots for NiO doped at low carbon and high carbon levels (5 x 5 µm$^2$). The solid lines represent metallic to insulating transition, the dotted line represent insulator to metallic transition.

It is important to note that the effect of carbon doping is not binary and in fact Figure 7 shows the gradual effect of the level of carbon doping in NiO. There is likely a threshold carbon doping to activate the switching mechanism and further increase in doping can be used to tune the CeRAM switching characteristics. Specifically, as shown in Figure 7, increase in the carbon doping reduces the operating voltage (SET and RESET) but increases the current in the semi-insulating state. We expect that the ability to tune such trade-offs will be extremely useful for engineering CeRAM for different applications [4].

Purposely adding carbon to TMOs and post-TMOs is not typical in the semiconductor industry. One example is the use of high-K HfO$_2$ as gate oxides [8], where carbon is known to cause increased leakage, and therefore efforts are typically directed at eliminating it [9]. In our case, as shown previously (Figure 7), carbon doping is identified with an increase in the conduction of the semi-insulating phase. Such semi-insulating behavior is essential for a proper operation of the CeRAM cell, in order to prevent hard breakdown by electro-forming seen in ReRAM or CBRAM.

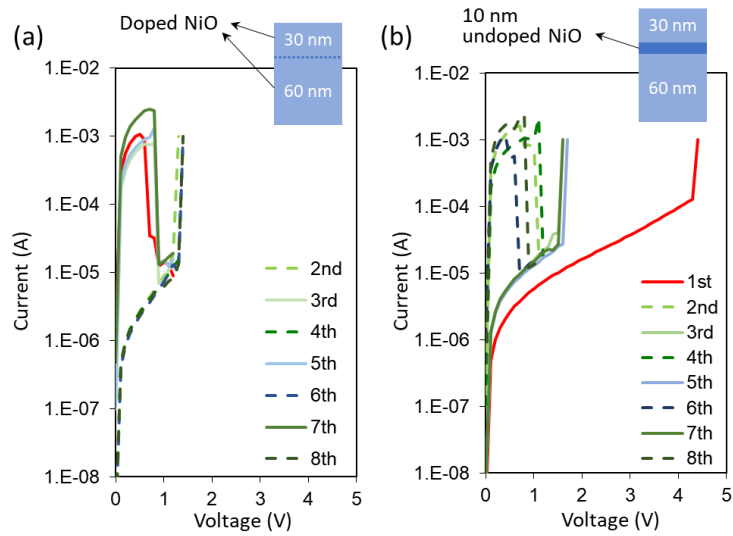

Figure 8: (a) Successive device measurement (voltage) sweeps of capacitor made of 30 nm carbon-doped NiO directly followed by 60nm of carbon-doped NiO (b) Corresponding successive device measurement of capacitor made of 30nm carbon-doped NiO, 10nm NiO followed by 60 nm of carbon-doped NiO. The inset shows a schematic of the device structure used in the measurements.

To further understand the effect of the carbon doping, we devised the experiment in Figure 8. In Figure 8a, we deposit a 90 nm thick carbon doped NiO CeRAM and in Figure 8b we add an additional 10 nm undoped NiO between two sections of 30 nm and 60 nm carbon doped NiO. In Figure 8a, we observe the expected virgin born-on behavior of CeRAM, and the device operation is as expected. However, the device with the undoped NiO inserted layer is virgin born-off due to large resistance from the undoped NiO layer, as shown in the 1st measurement (voltage) sweep. When a large voltage (4.5 V) is applied in this first sweep the undoped layer breaks down and the device characteristics are similar to CeRAM in subsequent sweeps. We postulate the undoped NiO undergoes a filament formation (like ReRAM and CBRAM) at high voltage, becomes electrically conducting and shorts the doped regions on either of its side. Once formed, the undoped region seems to play no active role in the device operation.

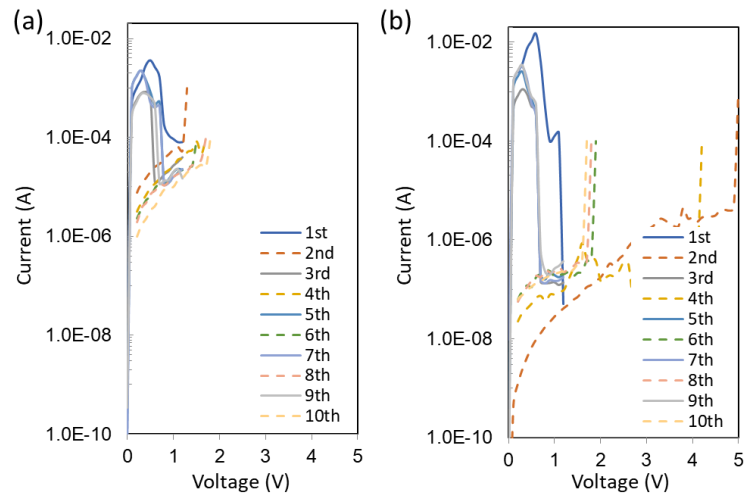

Figure 9: $YTiO_x$, both doped with C. Please note the difference in x-axis between the two figures. (a) Functioning CeRAM device and (b) Hard breakdown into ReRAM. The labels represent the sweep order.

Any oxide with sufficiently large electric field will undergo filament formation and eventual breakdown. In fact, if proper care is not taken during device fabrication and measurements, it is possible the CeRAM might undergo filament formation. We show this behavior in $YTiO_3$ devices fabricated on the same wafer. Though most of the devices show the expected CeRAM behavior shown in Figure 9a, we observe few devices shown in Figure 9b with different I-V characteristics. Both types of virgin devices show born-on behavior, but in Figure 9b a higher current and voltage is needed to switch the device which then experiences a deeper reset (to lower current) than experienced in Figure 9a. These devices in Figure 9b, now normal TMO capacitor, allows forming and standard ReRAM behavior. The excess current is most likely due to charge buildup at the interface, which is specific in our case to $YTiO_3$ and could be process or defect driven. This suggests that the CeRAM state can be broken and does differ from ReRAM filamentary devices, even if this difference is subtle.

Another point to notice from Figure 9 is that CeRAM devices exhibit larger leakage current, which we argue is due to different leakage mechanism in CeRAM versus ReRAM. The high resistance state in ReRAM is from a truly insulative material but in the CeRAM the state shows semi-insulating behavior. Similar results have been presented elsewhere [10, 6] that also suggest the high resistance state is not truly insulative but instead shows significant conductions likely from a lower band gap.

**Photoconductivity spectroscopy of NiO**: To determine the critical energy parameters of the electronic band structure of the carbon-doped NiO films we employed photoconductivity spectroscopy. The results of these measurements are illustrated in Figure 10 which shows photocurrent yield (i.e., photocurrent normalized to the incident photon flux) spectra collected in the visible-UV and near IR-visible part of the spectra in Figure 10a and Figure 10b, respectively.

The UV spectra were collected in the transient photocurrent mode [11] in the Si/SiO2/NiO capacitor (structure and the voltage biases are shown in the Figure 10a ) and reveal the photoconductivity threshold at 4.0 eV which is consistent with the earlier reported values for thin NiO films deposited by sputtering [12] (or by chemical vapor deposition [13]). This observation suggests that the "host" matrix corresponds to the stoichiometric NiO. There is additional density of electron states 'stretching' over ≈ 0.1 eV into the NiO gap. This observation is corroborated by the near-IR spectra (Figure 10b) which reveal additional spectral threshold of photoconductivity at significantly lower energy than that of the NiO. The photocurrent spectral curves collected using different low pass filters (Si and 780 nm) consistently yield the photoconductivity energy onset of 0.96 eV if fitted using indirect gap model as illustrated in Figure 10b . We suggest that this onset would likely correspond to the excitation of gap states introduced by carbon doping. Put together, we argue that the carbon doping introduces an additional set of states inside the original 4.0 eV gap, with the Fermi level inside this new manifold of states.

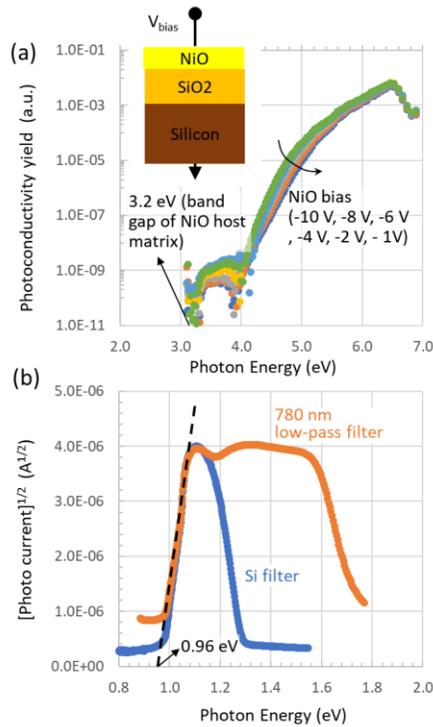

Figure 10: (a) Photoconductivity yield spectra in UV spectral range as measured on Si/SiO2/NiO structure (shown in the inset) under different negative bias voltages (from -1 to -10 Volt) applied to the top NiO electrode. The photocurrent onset at ≈4.0 eV corresponds to the bandgap of the "host" NiO matrix. (b) Near-IR photocurrent spectra of the same sample measured using several low-pass optical filters (silicon and 780 nm glass filters). The spectra shown as the (photo current)$^{1/2}$ vs. photon energy allow determination of a 0.96 eV wide indirect gap by linear fit of the near-threshold region as shown by the dashed line.

In the spectral range close to the Si substrate bandgap width (1.12 eV at 300 K) no normalization to the incident photon flux is applied because of strong change in the transparency of the Si crystal. The latter causes a peak-like structure near hv =1.1 eV.

**Band structure of undoped and doped NiO:** We reconcile our observations so far and propose the band structures for undoped and doped NiO in Figure 11. The addition of carbon causes back donation to the metal-ion, which is assumed to be responsible for the hybridization of orbitals that results in the states creating the small bandgap [14]. Based upon other evidence that the doping of NiO introduces holes [15], we argue that the new states are induced by carbon and live 0.96 eV above the valence band maximum (VBM). For "born-on" CeRAM, the Fermi energy lies within these created states and thus shows weak metallic behavior seen in Figure 4c.

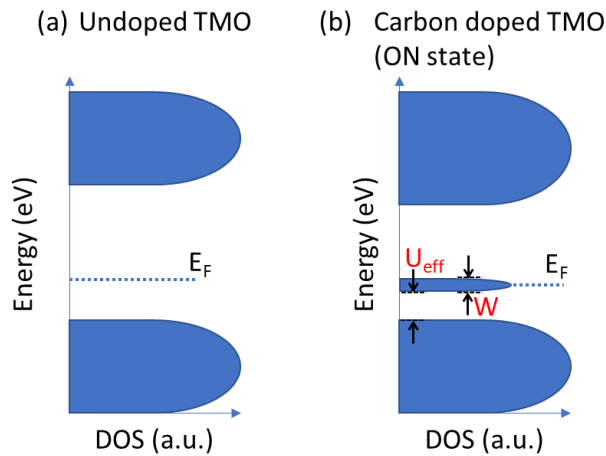

Figure 11: Band structure of an (a) undoped TMO, (b) a carbon doped TMO in ON state. The doped state has a narrow width "W" and should experience correlation effects.

The bandwidth of these new in-gap states is likely to be relatively narrow, coming from the overlap of the doped carbon states, which on average are far apart compared to the transition metal or O atoms, but much closer than the impurities in a typical doped semiconductor which has much lower doping levels. Due to the narrow band, these states are susceptible to strong correlation effects [16] (W ~< U, where U is an on-site Coulomb energy). Once in this regime, the system is likely to be unstable to small perturbations of either U or W, which may occur during the switching operation, though the precise details of how that occurs remain a subject of our research.

While 0.96 eV gap is small compared to the band gap of the host material, it is still very large compared to thermal energies (25 meV at room temperature) so none of the doped states in the new band should be thermally excited. This implies that the conduction of the material will be dominated by the new states inside the original band gap. This contrasts strongly with the case of a typical doped semiconductor where the impurity levels are near the valence or conduction bands and transport is dominated by thermally excited carriers (electrons or holes) into these bands. The carbon-doped metal oxides are instead similar to many doped Mott insulators (for example the cuprate high temperature superconductors) where the transport is much more exotic (e.g., having "bad metal" conductivity) and is dominated by the new states deep inside the gap. Such a "bad metal" behavior may be observed in our materials as well – see Figure 4c.

NiO is also a known Mott material [17]. The possibility of a Mott-like transition in the near-surface region at the positive electrode and NiO was first suggested by Inoue *et al* [18]. In unipolar ReRAM devices, filaments are observed to grow gradually, shorting the distance between anode and cathode (the cathode migrates to the anode electrode). High fields at the end of a filament could generate enough local electron density to cause a Mott-like transition as suggested in [18]. Forming-free unipolar devices, such as those in [6], could simply be due to carbon contamination during fabrication, which is not farfetched given the difficulty in detecting carbon at low levels. For CeRAM, we stabilize the system with purposeful carbon doping and make effort to keep it in the background matrix.

In typical undoped NiO the wide bandgap creates a highly resistive insulator which requires electroforming and filament formation resulting in ReRAM. The semi-insulating state of our doped compound is a different system that switches prior to the breakdown voltage of the undoped compound (as discussed in Figure 1).

4. Integration Results

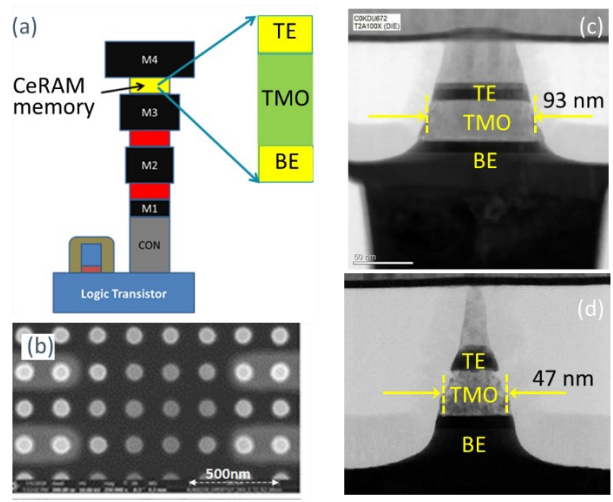

Figure 12: (a) Schematic drawing of CeRAM memory in series with logic transistor (1T1R bit cell) embedded in BEOL, (b) top-down SEM of array after pillar etch, (c) cross-sectional TEM showing critical dimension (CD) of ~93 nm and (d) ~47 nm devices.

To demonstrate scalability to and compatibility with advanced CMOS process nodes, CeRAM devices were fabricated to sub-50 nm dimensions using CMOS-compatible integration on 300 mm wafers. The integrated process includes a 65 nm technology node transistor at the front end, with CeRAM devices integrated into the back end of the line (BEOL) metal interconnects as shown in Figure 12a. Figure 12b shows a top view of an array of CeRAM devices, where the test chip includes arrays of devices with varying dimensions. Two transmission electron microscope (TEM) cross-sections of 93 nm and 47 nm CeRAM, are shown in Figure 12c and Figure 12d, respectively.

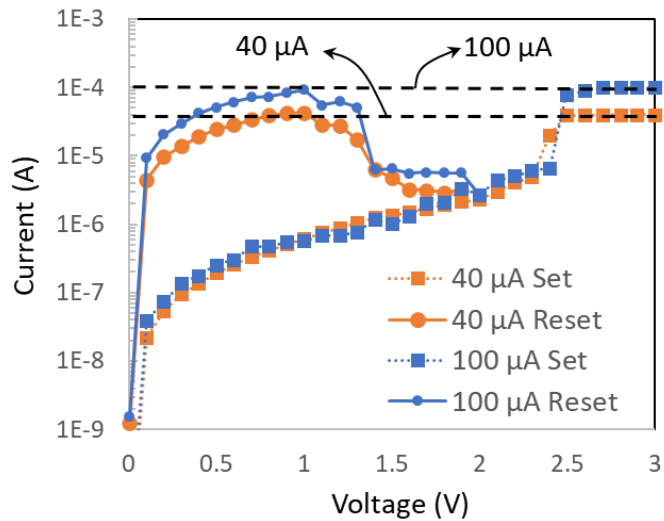

Figure 13: I-V curves (SET and RESET) for 93 nm devices demonstrate born-on cycling with compliance currents of 40 µA and 100 µA.

The DC I-V curves for 93 nm devices in a 1T1R configuration are shown in Figure 13. The smaller size of the devices compared to our large devices presented in the previous sections is reflected in lower switching currents (10s of µA vs. mA). Note that the switching voltages are slightly larger, which is due to voltage dropped across the series transistor, which was not optimized for these CeRAM devices. The actual CeRAM switching voltages are expected to be the same as the large devices shown in Section 2 and 3.

We use the series transistor to enable two distinct compliance currents of 40 μA and 100 μA. In both cases, after the compliance is achieved the devices are set in the ON state with corresponding different resistance. Note that the OFF conductance remains unchanged and indicates that the compliance current does not affect the film (such as forming a filament or altering a filament) and carrier transport is not trap limited. $V_{Set}$ and $V_{Reset}$ remain the same irrespective of the compliance. We also maintain a large memory window of >100x in these sub-100 nm devices.

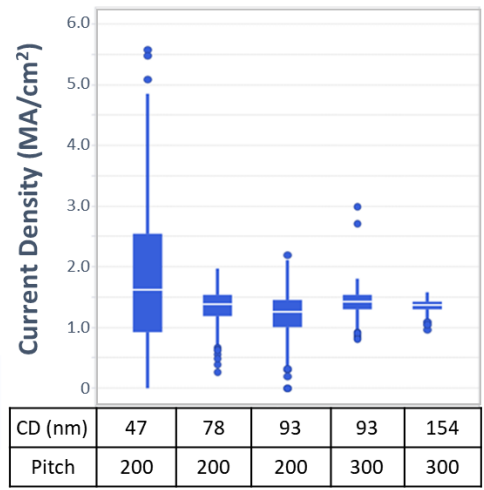

Figure 14: Boxplot of initial currents measured with different pillar sizes (47, 78, 93, 154 nm) and pitch (200 and 300 nm). The white line is the median, and the blue boxes on either side represent the 25 and 75 percentile of the data. If some devices are far beyond this range, they are represented by dots. CeRAM ON current shows area scaling. The increase in variation at smaller CD is due to integration window controls and not inherent to CeRAM.

ON currents shown in Figure 14 measured from multiple groups of devices with pillar sizes from 47 to 154 nm show area scaling of current. These currents are measured before switching and in the born-on virgin state. The constant current density indicates bulk conduction and conclusively shows an absence of filamentary conduction in the virgin state. Filamentary conduction typically manifests as current (not the current density) remaining constant with the area. Additionally, the constant current density indicates that when the devices are scaled to smaller devices, the power and current do scale proportional to area.

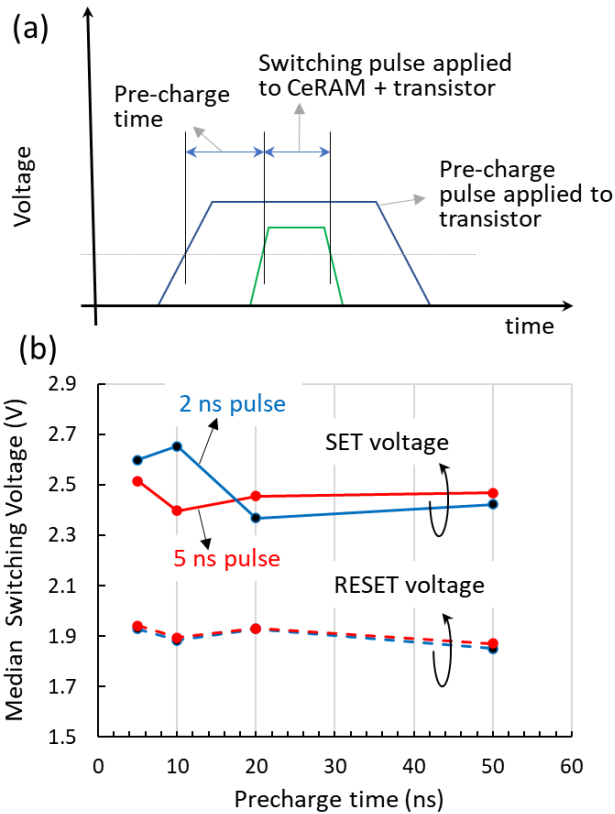

Figure 15: (a) The schematic to explain the switching pulse configuration to mimic commercial applications. The transistor gate is turned ON to pre-charge the interconnect and the device is switched on after a delay by application of the switching voltage to 1T1R. (b) Device read current at 0.3V after successful switching (SET and RESET) is plotted for different pulse widths down to 2 ns.

Figure 15a shows a typical pulsing operation of CeRAM device used in commercial applications. The transistor gate voltage is turned on for a "pre-charge time" before the switching voltage is applied to the CeRAM. In Figure 15b, we show the median SET and RESET switching voltage for 5 sub-100 nm CeRAM devices. The switching voltage does not change with the pre-charge time or the switching pulse width (2 ns vs 5 ns). Based on this data and other in process, we argue that (a) the achievable switching speed is well below 2 ns and (b) the switching is not energy or power driven. In filamentary devices such as ReRAM, the switching voltages drastically increase as the write pulse widths are reduced [19].

In Figure 16, we show that the read current for ON as well as OFF state shows almost no change as we heat the CeRAM devices up to 473 K (up to 24 hours), which is expected from our measurements on large devices and is a distinguishing trait of CeRAM – minmal degradation with temperature (and also consistent with Ref [6]).

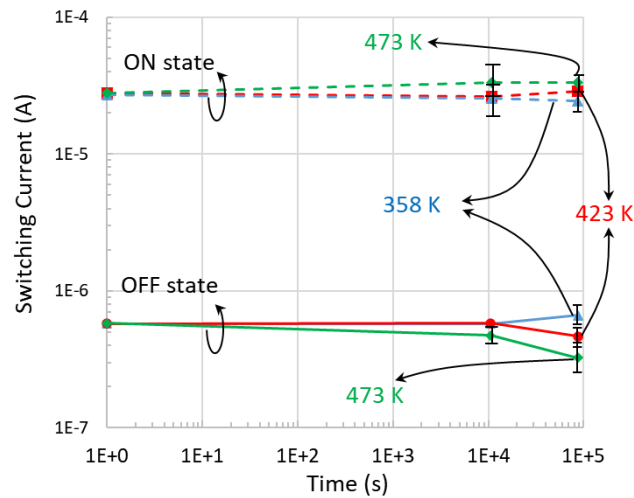

Figure 16: Data retention measured at 358 K to 473 K for 100 µA compliance current showed minimal degradation over 24 hours. Each data point includes 30 devices. The error bars for data points not shown are smaller than the symbols.

We use the pulse write method to program direct-wired 66-bit arrays and row-column addressable 1 kbit arrays, which have one access transistor for each cell. In Figure 17a and Figure 17b we show three operations on 66 bit and 1 kbit arrays respectively. The first operation RESETs all devices and read current is shown in red in Figure 17. The red color represents low READ current and indicates a successful RESET. The second operation SETs all devices and the READ current after SET is shown in green in Figure 17. The green color represents high current and indicates a successful SET. The minimum difference between low resistance state and the high resistance state for a successful operation is 10×. The final operation performs RESET on alternate devices and the resulting pattern looks like a checkerboard (alternate green and red squares). We can see that the 66-bit array show 100% percent functionality while the 1 kbit array shows 90% functionality. We further investigate 66-bit array by cycling it 1000 times. The cumulative READ currents for all 66 bits are shown in Figure 17c at intervals of 1, 10, 100, and 1000 cycles. For the first measurement of CeRAM technology, it is encouraging to see that 70% of the devices are alive after 1000 cycles.

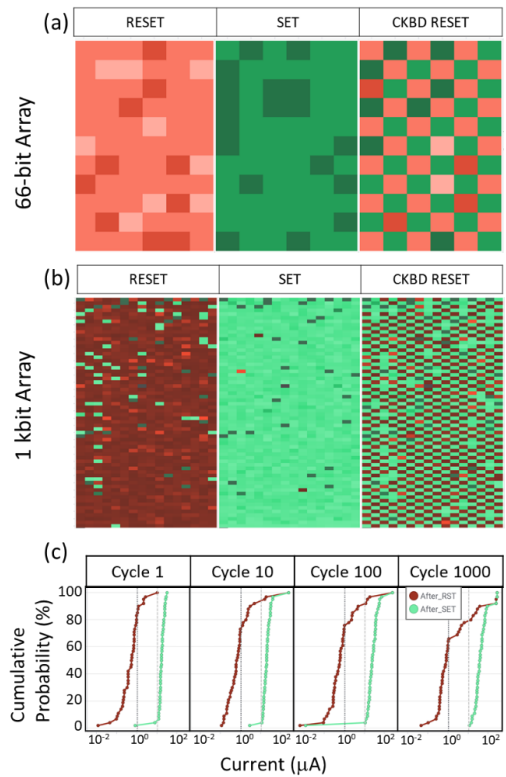

Figure 17: (a) 100% functional 66-bit mini-arrays after three operations and (b) 90% functional 1 kb arrays after three operations. The red color indicates low read current in high resistance state, whereas the green color indicates

high read current in the low resistance state. (c) 66-bit arrays of 65 nm (drawn dimension) devices were cycled, up to 1,000 times. ~70% of the devices successfully cycled.

## 5. Conclusion

A novel resistive switching device, CeRAM, that can be used in a variety of applications has been described. Carbon doping induces metal-like states universally across TMOs and post-TMOs and leads to better technological control of the device operation, which eliminates the randomness of other resistive memories caused by stochastic filamentary behavior. Further underscoring a universal effect, we have demonstrated CeRAM behavior with multiple electrode materials and with multiple deposition techniques (spin-on and PVD). An experimental explanation for the effect of carbon doping is evidenced by the photoconductivity results supporting the formation of a narrow band above the valence band that we hypothesize to Mott-Hubbard like correlation. Based on our experiments and model, we propose that the metal-insulator transitions in multiple carbon doped metal oxides shown here are true electronic transitions that could possibly be explained by the Mott-Hubbard approach.

The CeRAM device switching controlled by external current compliance allows setting variable resistance levels, which may be used to enable analog neuromorphic computing as well as to extend memory density, or as an additional adjustment for CMOS process variability. We have also shown that CeRAM devices can operate at extreme temperatures from 1.5 K to 423 K and show state retention up to 673 K for 1 hour bake. We have shown 47 nm diameter devices integrated with CMOS transistors on 300mm wafers, exhibiting fast write times (2 ns), high temperature retention (24 hours at 473 K), and 90% functional yield in 1kb arrays, which demonstrates the applicability of carbon doped CeRAM for advanced CMOS node non-volatile memory.